# High tunnel magnetoresistance and magnetism in metastable *bcc* $Co_{1-x}Mn_x$ - based magnetic tunnel junctions

Kazuma Kunimatsu,[1,2] Tufan Roy,[3] Jun Okabayashi,[4] Kelvin Elphick,[5] Tomoki Tsuchiya,[6,7] Tomohiro Ichinose,[2] Masahito Tsujikawa,[3,7] Atsufumi Hirohata,[5] Masafumi Shirai,[3,7,6] and Shigemi Mizukami[2,7,6,*]

[1]*Department of Applied Physics, Graduate School of Engineering, Tohoku University, Sendai 980-8579, Japan*

[2]*WPI Advanced Institute for Materials Research, Tohoku University, Katahira 2-1-1, Sendai 980-8577, Japan*

[3]*Research Institute of Electrical Communication, Tohoku University, Sendai 980-8577, Japan*

[4]*Research Center for Spectrochemistry, The University of Tokyo, Tokyo 113-0033, Japan*

[5]*Department of Electronic Engineering, The University of York, York YO10 5DD, England*

[6]*Center for Science and Innovation in Spintronics (CSIS), Core Research Cluster (CRC), Tohoku University, Sendai 980-8577, Japan*

[7]*Center for Spintronics Research Network (CSRN), Tohoku University, Sendai 980-8577, Japan*

[*] shigemi.mizukami.a7@tohoku.ac.jp

Co-rich $Co_{1-x}Mn_x$ alloys have hcp or fcc disordered phases and those ferromagnetic orderings are significantly deteriorated with increasing Mn concentration $x$ in bulk. On the other hand, those metastable *bcc* phases show properties attractive to spintronics, *e.g.,* high tunnel magnetoresistance (TMR) ratio of more than 200% (600%) at 300 K (10 K) in magnetic tunnel junctions (MTJs) with the $x$ = 0.25 *bcc* alloy electrodes [Kunimatsu *et al.,* Appl. Phys. Express 13, 083007 (2020)]. Here, we report systematic study of structure and magnetism for epitaxial thin films as well as the TMR effect in MgO(001)-barrier MTJs with electrodes comprising those *bcc* films. The single phase *bcc* $Co_{1-x}Mn_x$(001) films were pseudomorphically grown on Cr(001) for $0.14 < x < 0.50$ with a sputtering technique. The magnetization was larger than that of pure Co for $x$ = 0.14– 0.25 and deceased with further increasing $x$. This behavior mainly stemmed from the composition dependence of magnetic moment of Mn that exceeded 2 $\mu_B$ at the maximum, unveiled by X-ray magnetic circular dichroism. Correspondingly, within the range of $0.25 < x < 0.37$, the TMR ratio decreased from 620% (229%) to 450% (194%) at 10 K (300 K) as $x$ increased. We discussed the relationship between the magnetism and high TMR ratio with different $x$ with the aid of the *ab-initio* band structure calculations.

## I. INTRODUCTION

A magnetic tunnel junction (MTJ) exhibits tunnel magnetoresistance (TMR) effect [1– 3], and is utilized in various storage system [4, 5] and non-volatile magnetoresistive random access memories (MRAM) [6]. Currently, FeCo(B) alloy electrode and MgO barrier are used for MTJs as the standard materials [7]. Those MTJs are known to exhibit a large tunnel magnetoresistance (TMR) effect, up to 604% at room temperature (RT) after device annealing at relatively high temperature, such as 550°C [8]. Recent interest in artificial intelligence technology has led to proposals for advanced spintronic applications, such as MTJ-based non-von-Neumann computing circuits, one of those requirements is much higher TMR ratio, typically 500-1000% at RT [9–11]. Therefore, it is of interest from both a fundamental and technological perspective to explore the physics of MTJs that exhibit such huge TMR ratios at room temperature.

Most of the previous works focused on *bcc* FeCo(B) [7, 8, 12–20] and $Co_2$-based full Heusler alloys [21–25] for magnetic electrodes to obtain large TMR effects. This is due to the theoretical prediction of the coherent tunneling for MgO-based MTJs with Fe(001) or B2 ordered FeCo(001) electrodes [26, 27] and because of the half-metallic electronic structures for $Co_2$-based full Heusler alloys [28]. In fact, the experimental TMR ratios exceed 1000% at low temperature in MgO-based MTJs with FeCo(B)(001) electrodes [8] and $Co_2MnSi$(001) Heusler alloy electrodes [24], whereas those values were significantly reduced at RT. Hence further research on materials and physics for MTJs is crucial to achieve such huge TMR effect at RT.

The huge TMR ratio based on the coherent tunneling mechanism has also been predicted in MgO-based MTJs with *bcc* Co(001) electrodes, discussed to be higher than those with Fe(001) electrodes [29]. Indeed, Yuasa *et al.* reported that MgO-based MTJs with the ultrathin *bcc* Co(001) electrodes exhibited the TMR ratio of 410% at RT and 507% at LT [30]. However there were no other experimental reports on the high TMR effect in MgO-based MTJs with *bcc* Co(001) electrodes and there are some debates [15, 16]. Generally, it is difficult to obtain nm-thick and single-phase *bcc* Co films [14, 16, 30–34], because *bcc* Co is not a true metastable phase of bulk Co, as theoretically suggested [35]. Thus the research on MgO-based MTJs with other *bcc* Co alloy electrodes would shed light on this issue and may help to pursuit advanced materials for MTJs.

One candidate is the metastable *bcc* $Co_{1-x}Mn_x$ [36–41]. In bulk of Co-rich $Co_{1-x}Mn_x$ binary, fcc or hcp disordered phase is thermodynamically stable [42–45]. For these phases, magnetizations are remarkably reduced with increasing the Mn concentration $x$, then magnetic phase transition occurs into antiferromagnetic ordering at $x = 0.3–0.4$. However, the magnetism of the metastable *bcc* $Co_{1-x}Mn_x$ is quite distinct at $x = 0.2 - 0.4$ from those of fcc/hcp phases. This metastable *bcc* phase was first reported as films which were grown on GaAs(001) with molecular beam epitaxy (MBE) [36, 37]. A critical thickness, at which the *bcc* structure relaxes to the others, depends on $x$ and was extended up to $\sim$ 55 nm for approximately $x = 0.25$ [37]. The *bcc* films have also been grown on MgO(001) with MBE, and their X-ray magnetic circular dichroism indicated that the magnetic moment of Co is ferromagnetically coupled to that of Mn and is maximized at approximately $x = 0.24$ [38]. The Curie temperatures $T_c$ for the *bcc* films with $x > 0.33$ have been investigated [39].

We reported 10-nm-thick *bcc* films with $x = 0.25$ fabricated by sputtering whose saturation magnetization, $\sim$ 1.6 MA/m, was slightly larger than that of fcc or hcp Co, $\sim$ 1.4 MA/m [40]. In addition, we reported, for the first time, an observation of the high TMR ratio of approximately 200–250% at RT, and of greater than 600% at low temperature in the $Co_3Mn$/MgO/$Co_3Mn$(001) MTJs [41]. This low temperature value was slightly higher than the above-mentioned value in MgO-based MTJs with the ultrathin *bcc* Co(001) electrodes [30].

In this article, we systematically study structures and magnetism of the metastable *bcc* $Co_{1-x}Mn_x$ disordered alloys epitaxial thin films as well as the TMR effect in those MgO(001)barrier MTJs with different $x$ to deeply understand the physics related to the TMR effect in the MTJs. In particular, we employed the X-ray magnetic circular dichroism (XMCD) to probe microscopic magnetism, and we discuss relationship between the magnetism and the TMR effect, based on the *ab-initio* band structure calculation of the bulk alloys.

## II. EXPERIMENTAL METHOD AND COMPUTATIONAL DETAILS

All samples were deposited on MgO(001) single crystal substrates using a magnetron sputtering under a base pressure of below $2\times10^{-7}$ Pa, similar to the samples in the previous reports [40, 41]. We prepared the sample of MgO(001)/Cr(40)/$Co_{1-x}Mn_x$(10)/MgO(2.4) ($x$ = 1.00, 0.14, 0.25, 0.34, and 0.50) for characterization of structure and magnetic properties. The MTJs stack prepared was MgO(001)/Cr(40)/$Co_{1-x}Mn_x$(10)/MgO(2.4)/$Co_{1-x}Mn_x$(4)/$Co_{0.75}Fe_{0.25}$(1.5)/$Mn_3Ir$(10)/Ru(5) ($x$ = 0.14, 0.17, 0.25, 0.34, and 0.37) (thickness in nm). Cr(001) was used to obtain atomically flat surface, as similarly employed before [46– 48]. The nominal composition $x$ was varied with a co-sputtering using elemental targets of Co and Mn. Here, a $Co_{0.75}Fe_{0.25}$ layer was deposited on the top $Co_{1-x}Mn_x$ layer to enhance an exchange bias [41]. All layers were deposited at RT. An in-situ annealing was carried out at 700°C and 200°C after the deposition of Cr and $Co_{1-x}Mn_x$, respectively. The MTJ multilayer films were patterned with a standard ultraviolet photo-lithography and Ar ion milling. The thirty six MTJs were fabricated on the substrate with the junction areas of 10×10, 20×20, and 30×30 $\mu m^2$. The MTJs were annealed at 325°C for one hour in a vacuum furnace with applied magnetic field of 0.5 T.

The film structures were characterized using x-ray diffraction (XRD) using Cu $K_\alpha$ radiation. Magnetization measurements were performed using a vibrating sample magnetometer (VSM). X-ray absorption spectroscopy (XAS) and XMCD measurements were performed at BL-7A, High-energy accelerator research organization, Photon Factory (KEK-PF), Japan. The XAS spectra with different helicities were obtained by switching the magnetic fields in the parallel and antiparallel directions along the incident beam. The total-photoelectron-yield mode was adopted by directly detecting the sample current. XAS and XMCD measurements were performed at room temperature. A magnetic field of ±1 T, produced using an electric magnet, was applied along the direction of incident polarized soft X-rays. In order to detect the in-plane magnetic signals, the sample surface normal direction was rotated 60 deg from incident beam and magnetic field [48–50]. Since the probing depth of photoelectrons was approximately 5 nm in the TM L-edge photon energy regions. Transmission electron microscopy (TEM) measurements were carried out for cross-sectional structural analysis for the MTJs. Electrical transport properties were measured by a four-probe method using a physical property measurement system (PPMS) with different temperatures and an applied voltage of around 1 mV.

*ab-initio* calculations were carried out using the spin-polarized-relativistic Korringa-Kohn-Rostoker (SPRKKR) method, as implemented in the SPR-KKR program package [51]. The effect of substitutional disorder has been considered by coherent potential approximation. For the exchange correlation functional, the generalized gradient approximation, as parameterized by Perdew, Burke, and Ernzerhof (PBE) [52], was used. An angular expansion of up to $l_{max} = 3$ has been considered for each atom. We have used 1183 irreducible *k*-points for the Brillouin zone integrations.

## III. EXPERIMENTAL RESULTS

Figure 1(a) shows the out-of-plane $2\theta$–$\omega$ XRD patterns for the $Co_{1-x}Mn_x$ films. The Co film, *i.e.*, $x = 0$, showed the diffraction peak at ∼ 75 deg in addition to the (002) peak from the Cr buffer and the (002) peak from the MgO substrate, which may be attributed to that from fcc (220) or hcp (11-20) of Co. For $0.14 < x < 0.50$, the $Co_{1-x}Mn_x$ films show the diffraction peak much closer to the Cr (002) peaks and there found no other peaks stemming from the films. The diffraction peaks of $Co_{1-x}Mn_x$ systematically shift with the Mn concentration $x$, and the peak positions are close to those for the (002) diffraction peaks expected from the lattice constant of *bcc* $Co_{1-x}Mn_x$. Thus the XRD data suggests that the formation of the metastable *bcc* $Co_{1-x}Mn_x$(001) alloys on Cr(001) for the present range of $x$. Figures 1(b)-1(d) show reciprocal space mapping of the films with typical $x$. The diffraction spots from (101) of Cr are clearly visible and those of *bcc* $Co_{1-x}Mn_x$ partially overlap with Cr (101), as observed in Figs. 1(b) and 1(c). The vertical positions of (101) of *bcc* $Co_{1-x}Mn_x$ shift along the MgO [001] direction, namely a film normal, with different $x$, whereas those horizontal positions seems to be identical to the (101) of Cr. Figure 1(d) indicates that (101) of *bcc* $Co_{1-x}Mn_x$ almost overlaps with the (101) of Cr. These data of the reciprocal mapping suggest that *bcc* $Co_{1-x}Mn_x$(001) is grown on Cr(001) in a pseudomorphic fashion with matching their in-plane lattice to that of Cr(001).

Figure 2 shows the out-of-plane lattice constants $a_{[001]}$ evaluated from the (002) diffraction peaks of $Co_{1-x}Mn_x$. The out-of-plane lattice constants $a_{[001]}$ linearly change with varying the Mn concentrations $x$. We assume that the $Co_{1-x}Mn_x$ films in this study have a *bcc* cell with slight tetragonal distortion, then we evaluated the lattice constant of *bcc* cell $a_c$ without such tetragonal distortion. For this evaluation, we assume that the unit cell volume is preserved against tetragonal distortion in each $x$ and the in-plane lattice constant of the $Co_{1-x}Mn_x$ film is identical to that of Cr, 0.288 nm. The calculated lattice constants $a_c$ in the *bcc* cell without tetragonal distortions are shown in Fig. 2. The lattice constant of *bcc* Co extrapolated from the data of $a_c$ vs $x$ is close to the literature values, 0.282–0.283 nm [31, 32]. Here, the $a_{[001]}$ values are quite close to that of $a_c$ or the lattice constant of Cr at $x$ of around 0.25–0.34, indicating that tensile or compressive strain existing in the $Co_{1-x}Mn_x$ films below or above $x \sim 0.3$, respectively. The (002) diffraction peaks for the $Co_{1-x}Mn_x$ films show the minimum for the line-broadening at $x$ of around 0.25, and this could correlate with the strain state in the $Co_{1-x}Mn_x$ films depending on $x$.

Figure 3(a) shows in-plane magnetization hysteresis curves for the $Co_{1-x}Mn_x$ films with different Mn concentration $x$. The Co film exhibits relatively large coercive force and its magnetization saturated at the magnetic field of around 0.4 T. On the other hand, the $Co_{1-x}Mn_x$ films show relatively smaller coercive force and those magnetizations saturate under relatively low magnetic field. The $x$ = 0.5 films shows negligible magnetization. Figure 3(b) displays the saturation magnetizations $M_s$ as a function of $x$. The $M_s$ values lie in between 1.3–1.7 MA/m for $x$ = 0.14–0.34 and those are comparable to or slightly higher than that of the Co film, ∼ 1.4 MA/m. The enhanced magnetization observed in this study is qualitatively consistent with the past report of Snow *et al.*, where they have suggested that *bcc* $Co_{1-x}Mn_x$ with $x$ of around 0.25 has the magnetic moment of Co larger than that in pure Co [38].

Figure 4 displays the typical data of the XAS and XMCD spectra for Co and Mn $L$-edges measured in the $x$ = 0.25 film. The XAS spectra were normalized to one at the post-absorption edge. Almost similar metallic line shapes were detected for both Co and Mn edges. The element specific magnetization curves at each Co and Mn $L_3$-edge XMCD in Fig. 4(c) also exhibit similar shapes as in-plane easy axis due to the strong exchange coupling between Co and Mn. Intensity ratios between Co and Mn are also changed systematically. The similar XAS and XMCD spectra were obtained for the films with other $x$, except for the $x$ = 0.5 film. We cannot detect the XMCD spectrum for the $x$ =0.5 films within the instrumental uncertainty, being consistent with negligible magnetization detected by VSM measurements in this films (Fig. 3).

Figure 5 shows the magnetic moments estimated from the XMCD measurements for the $Co_{1-x}Mn_x$ films. Magneto-optical sum rules were adopted to deduce spin and orbital magnetic moments using the spectral integrals of each XAS and XMCD. The total magnetic moments $M_{\mathrm{tot}}$ evaluated from the sum rules are in good agreement with those obtained from the magnetization measurements [Fig. 5(a)]. Note that $M_{\mathrm{tot}}$ for fcc $Co_{1-x}Mn_x$ is much smaller, ∼ 0.5 $\mu_B$ at $x$ = 0.25 and negligible values at $x$ = 0.35 [43]. As seen in Fig. 5(b), the magnetic moment of Co ($M_{\mathrm{Co}}$) for the $x$ = 0.14 films is 1.91 $\mu_B$, being slightly larger than that for the Co film, 1.75 $\mu_B$. Then, the values of $M_{\mathrm{Co}}$ only slightly decrease with increasing $x$. On the other hand, the magnetic moments of Mn ($M_{\mathrm{Mn}}$) exhibit relatively large change with $x$ [Fig. 5(c)]. The $M_{\mathrm{Mn}}$ values are in the range of 1.6–2.3 $\mu_B$ for $x$=0.14–0.34. The enhanced $M_{\mathrm{Co}}$ qualitatively agrees with the past report $M_{\mathrm{Co}}$ = 2.38 $\mu_B$ at for the $x$ = 0.24 films; however the enhancement observed in this study is much lower.

Figure 6 shows the typical cross-sectional TEM images of the MTJs with $x$ = 0.34. The data indicates the lattice coherence from the bottom $Co_{1-x}Mn_x$ to the top $Co_{1-x}Mn_x$ layer through the MgO barrier. Such lattice coherence is prerequisite for the coherent tunneling transport. The TEM data indicated that the in-plane lattice of the top $Co_{1-x}Mn_x$ layer on MgO(001) is slightly large as compared with that of the bottom. This may result from the tensile strain of the epitaxial growth of the top $Co_{1-x}Mn_x$ layer on MgO(001). The above-mentioned structural properties were similarly observed for the MTJs with the electrodes with $x$ = 0.75. The strained lattice may be reasonable to the observation in Figs. 1(c) and 1(d). More detailed structural analysis will be discussed elsewhere.

Figures 7(a), 7(b), and 7(c) show the MR curves measured at 300 K, 100 K and 10 K, respectively, in the MTJs with the $Co_{1-x}Mn_x$ electrodes with different $x$. Here, MR is defined as $[R(H) - R_p]/R_p \times 100$ with the resistance $R$ and the one at the parallel state $R_p$. The hysteresis of the MR curves reflects the differences

in the switching field of the magnetization for the top and bottom $Co_{1-x}Mn_x$ electrodes. The low or high resistance corresponds to the parallel or antiparallel state of the magnetizations, respectively. In the data measured at 300 K, the resistance changes near the field of ±10 mT are attributed to the magnetization reversal for the bottom $Co_{1-x}Mn_x$ layer, a free layer. The resistance changes at the field of ~ −20 mT and ~ −10 mT stem from the magnetization reversal of the top $Co_{1-x}Mn_x$ layer which was pinned by the exchange biased with IrMn via the CoFe layer. Here, the CoFe layer helps to increase the exchange bias, and no clear plateaus corresponding to the antiparallel state of the magnetizations were observed at 300 K, if without CoFe [41]. Even though the CoFe layer was used, the MTJs with the $x$ = 0.14 and 0.17 film electrodes showed no clear resistance plateaus corresponding to the antiparallel state of the magnetizations. This may partially be caused by the increase in the coercive force for the bottom $Co_{1-x}Mn_x$ layer, which became comparable to the exchange bias shift after the annealing of the MTJs at 325°C. Therefore, we discuss only the properties for the MTJs with $x$ = 0.25, 0.34, and 0.37 in this study.

Figure 8 displays the TMR ratio as a function of the Mn concentration $x$ with different temperatures. The TMR ratio at 10 K is 620% for $x$ = 0.25 and the value systematically decreases to 450% for $x$ = 0.37 as $x$ increases. The TMR ratios at RT are around 200% and those composition dependences are relatively weaker than that at 10 K. The values of the TMR ratio are comparable to the typical TMR ratios reported in the fully epitaxial MgO(001) barrier MTJs with FeCo(001) alloys electrodes prepared with MBE, *e.g.,* 250-500% at 20 K and 150-300% at 300 K [15, 16]. We evaluate the tunneling spin polarization $P_{\text{eff}}$ of the $Co_{1-x}Mn_x$(001) from the TMR ratio at 10 K using the Julliere's formula [1]:

$$\text{TMR ratio} = \frac{2P_{\text{eff}}^2}{1 - P_{\text{eff}}^2} \times 100.$$

The evaluated values are shown in Fig. 8(b) as a function of the Mn concentration $x$. The values are around 0.85 and show only slight decreases with increasing $x$. The highest value is 0.87 evaluated at $x$ = 0.25. Note that the values in Fig. 8 are similar to the tunneling spin polarization of 0.85 evaluated in CoFe(001)/MgO(001)/Al junctions under superconducting state of the Al electrode [13].

**IV. DISCUSSION**

For the coherent tunneling in MgO(001)-barrier MTJs, the TMR ratio is predominantly determined by the spin polarization at the Fermi energy for the energy band with a $\Delta_1$ symmetry for magnetic electrodes due to the symmetry filtering effect of MgO(001) [26, 27]. To understand the composition dependence of the large TMR effect in this study from the view points of bulk electronic structure of the electrode alloys, we performed the *ab-initio* calculations of the bulk band structures for *bcc* $Co_{1-x}Mn_x$ disordered alloys with various compositions.

Figure 9 shows the band dispersion along the [001] axis with different Mn concentrations $x$, and Fig. 10 shows the corresponding density-of-states for the alloys. In these calculations, we assumed the experimental out-of-plane and in-plane lattice constants and ferromagnetically coupled Co and Mn. In those figures, we also show the data for the *bcc* Co ($a$ = 0.282 nm), for reference. The Fermi energy is lowering as $x$ increases as expected from reducing the total electron numbers and the band structures become broader with increasing $x$ due to the substitutionary disorders. The shifts of the Fermi energy are also clearly seen in Fig. 10. For the minority spin state in all $x$ shown here, the $d$ orbitals for Co dominantly contribute to the electronic states at the Fermi energy and that for Mn does at around 2 eV higher than the Fermi energy, due to the strong exchange splitting at Mn site. On the other hand, for the majority spin state, the electronic states at the Fermi level gradually increase with increasing $x$, which are attributed to the $d$ orbitals for Mn.

Regarding the band dispersion for *bcc* Co, the Fermi energy is located across the $\Delta_1$ band in the majority spin state [Fig. 9(a)] whereas the Fermi energy lies just below the bottom of the $\Delta_1$ band in the minority spin state [Fig. 9(g)]. Thus, the electrons in the $\Delta_1$ bands are fully-spin polarized and the tunneling transmission amplitudes between the $\Delta_1$ bands for the antiparallel state of magnetizations are negligible, leading the huge TMR ratios [29, 30]. When the Mn concentration $x$ increases, we can see the broadening of the bottom of the $\Delta_1$ band at the Γ point for the minority spin state, which induces the finite state at the Fermi energy [Figs. 9(g)-9(l)]. Thus this disorder effect may deteriorate the tunneling spin polarization of the $\Delta_1$ band and decrease the TMR ratio with increasing $x$, as discussed in MgO(001)-barrier MTJs with the FeCo(001) electrode [16]. These features qualitatively agree with the composition dependence of $P_{\mathrm{eff}}$ [Fig. 8(b)].

The influence of the $\Delta_5$ bands on the TMR effect has been discussed [26]. The transmission amplitudes of the electron tunneling with the $\Delta_5$ symmetry are negligibly smaller than that with the $\Delta_1$ symmetry for a thick MgO barrier; thus the effect of the $\Delta_5$ band on the TMR effect is not dominant in fully coherent tunneling predicted from the first-principles calculation [26]. On the other hand, it is likely considered that non-ideal atomic structure at interfaces of MgO mixes the $\Delta_5$ and $\Delta_1$ symmetries of electron states, leading decreases in the TMR ratio from the values expected in the coherent tunneling [53]. In our case, at $x = 0.25$, the Fermi energy is close to the top of the $\Delta_5$ band in the majority spin state [Fig. 9(c)], then further increases in the state with the $\Delta_5$ symmetry can be observed with increasing $x$ [Figs. 9(d) and 9(e)]. Hence this increase of the $\Delta_5$ states in the majority spin state could be also related to the composition dependence of $P_{\mathrm{eff}}$ [Fig. 8(b)].

On the other hand, when we compare the experimental magnetic moments of Mn with those obtained from the band structure calculation, we are aware of non-negligible differences in them, as shown in Fig. 5. The differences between them systematically increase as increasing the Mn concentration $x$ in $Co_{1-x}Mn_x$, so this might be caused by small amount of antiferromagnetic and atomic scaled Mn clusters, as discussed previously [38]. Very small amount of such atomic clusters, a few %, in the films may be enough to explain the magnetic moment reduction of Mn observed in the experiment. Such small amount of the Mn clusters may not significantly change the global electronic structure shown in Figs. 9 and 10, whereas those may have an impact on the spin-dependent tunneling when those exist near the interface of MgO. Thus, we cannot rule out the possibility that the slight reduction of the tunneling spin polarization $P_{\mathrm{eff}}$ is related to the change in the magnetic moment of Mn observed experimentally, rather than the change in the band structure shown in Fig. 9. Experimental characterization in atomic level is necessary for further understanding.

**V. SUMMARY**

We systematically studied structure and magnetism for epitaxial thin films of the metastable *bcc* $Co_{1-x}Mn_x$ as well as the TMR effect in MgO(001)-barrier MTJs with electrodes comprising those films. The XRD analysis clarified that the 10-nm-thick and single phase *bcc* $Co_{1-x}Mn_x$(001) films were grown on Cr(001) in pseudomorphic fashion for $0.14 < x < 0.50$. From the XMCD analysis, the magnetic moment of Mn of ~ 2 $\mu_B$ was evaluated at $x = 0.14$ and those decreased with increasing $x$. The TEM analysis indicated the coherent epitaxial growth of the MTJs. The TMR ratio decreased with increasing $x$ within the range of $0.25 < x < 0.37$, from 620% (229%) to 450% (194%) at 10 K (300 K). We discussed the relationship between the high TMR ratio observed and the spin polarization of the $\Delta_1$ band dependent on the substitutionary disorder of the alloy that was predicted from the *ab-initio* calculations. In addition, we found the large difference in the magnetic moment for Mn obtained from the experiments and first-principles calculations, whose origins and effects on the TMR effect were also discussed.

## ACKNOWLEDGMENTS

We would like to thank Y. Kondo for his assistance. This work was partially supported by JST CREST (No. JPMJCR17J5). Parts of the synchrotron radiation experiments were performed under the approval of the Photon Factory Program Advisory Committee, KEK (No. 2019G028).

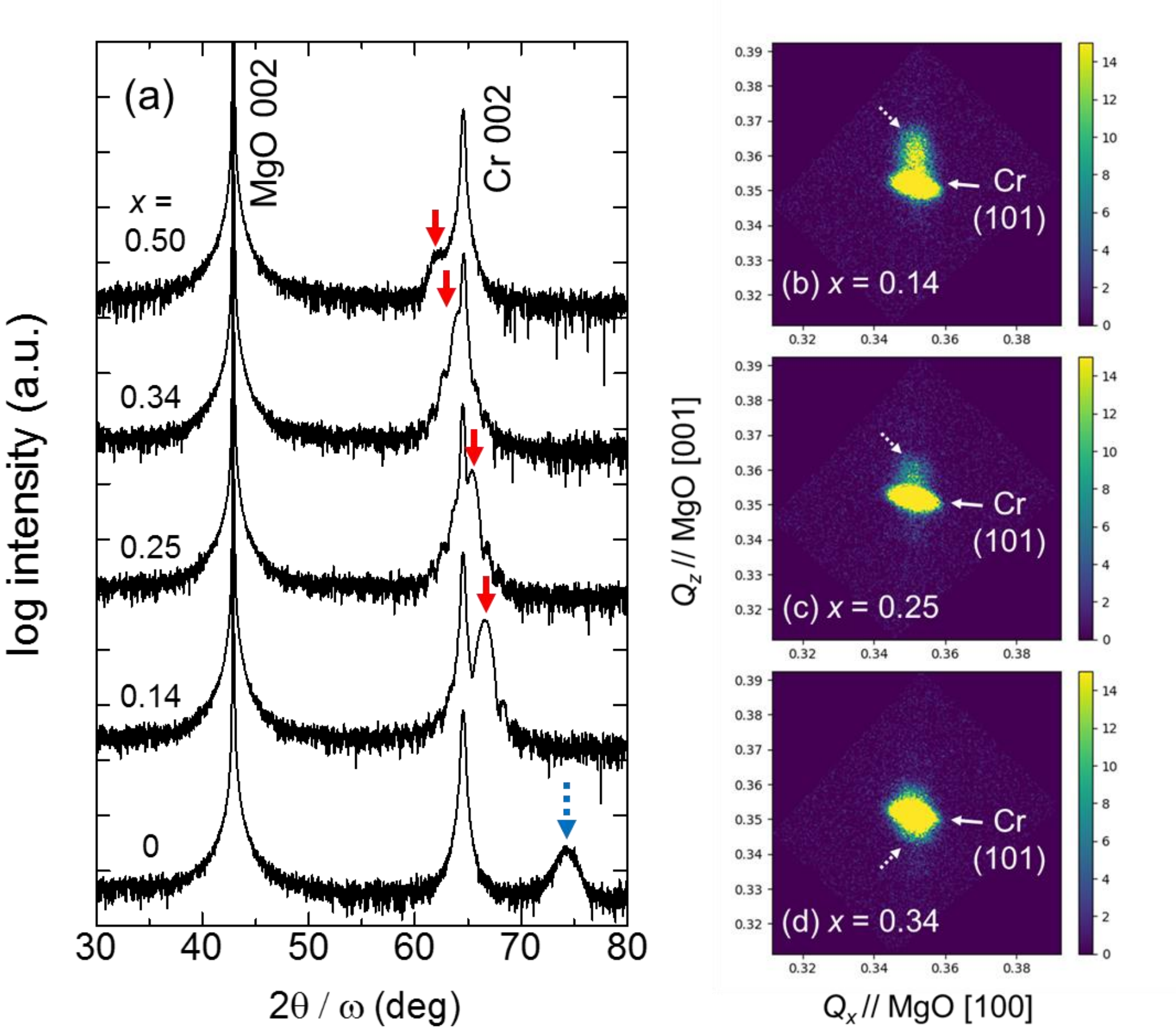

FIG. 1. (a) Out-of-plane $2\theta$–$\omega$ XRD patterns for the $Co_{1-x}Mn_x$ films. Arrows denote the diffraction peaks attributed to those of *bcc* $Co_{1-x}Mn_x$ (002), and dotted arrow denotes that of (11-20) for hcp Co or (202) for fcc Co. The reciprocal space mapping of the diffraction from (101) for *bcc* Cr and for *bcc* $Co_{1-x}Mn_x$ at (b) $x$ = 0.14, (c) 0.25, and (d) 0.34. Dotted arrows denote (101) for *bcc* $Co_{1-x}Mn_x$.

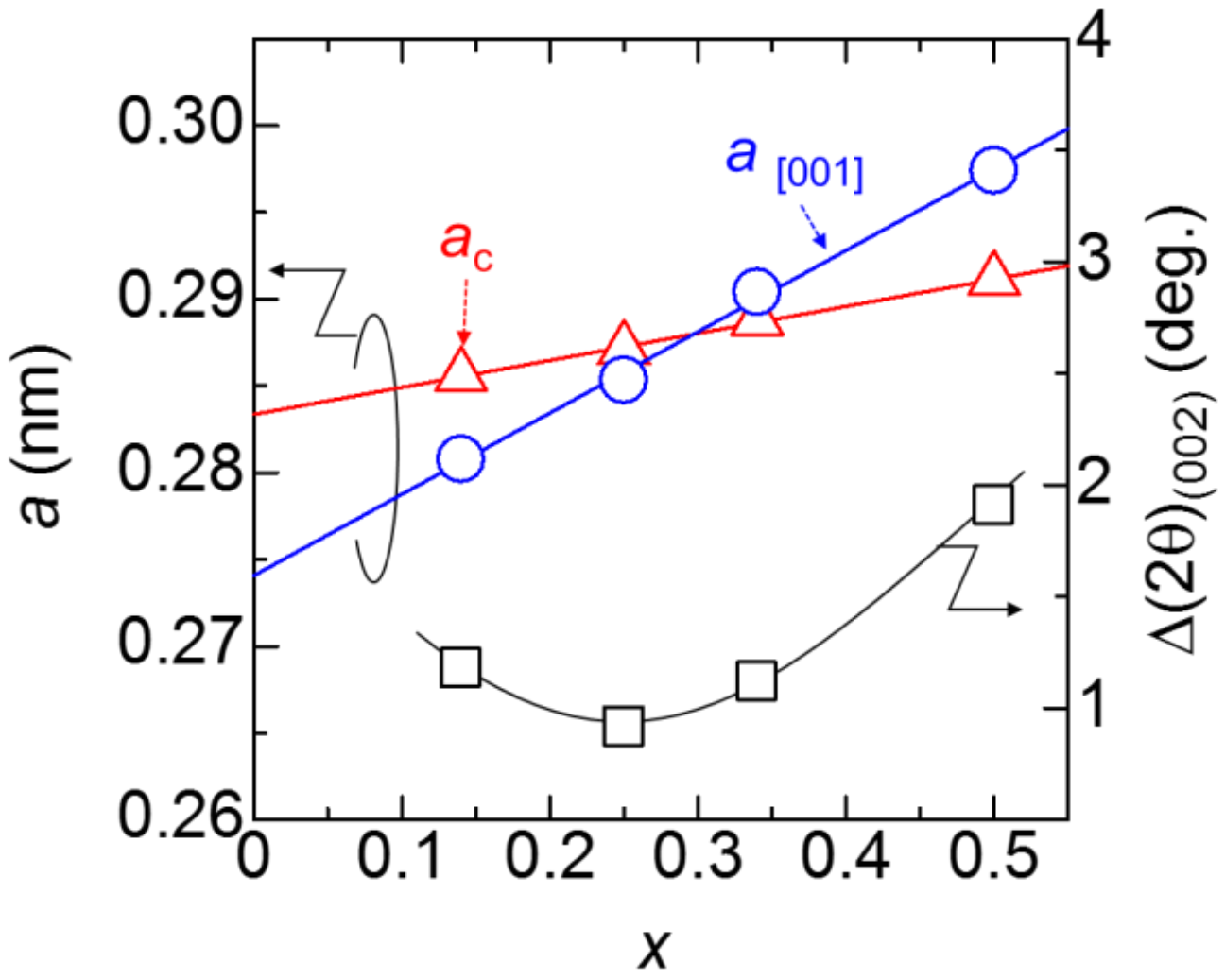

FIG. 2. Mn concentration $x$ dependence of the lattice constants $a_{[001]}$, $a_c$, and the full-width at half maximum of (002) diffraction peaks $\Delta(2\theta)_{(002)}$ for *bcc* $Co_{1-x}Mn_x$. Out-of-plane lattice constants $a_{[001]}$ was evaluated from the out-of-plane XRD. The lattice constants $a_c$ supposed for *bcc* $Co_{1-x}Mn_x$ without a tetragonal distortion were calculated from $a_{[001]}$ with fixing unit cell volumes and assuming the in-plane lattice constant identical to that of Cr (0.288 nm) at each $x$. Lines are the linear fits and curve is a visual guide.

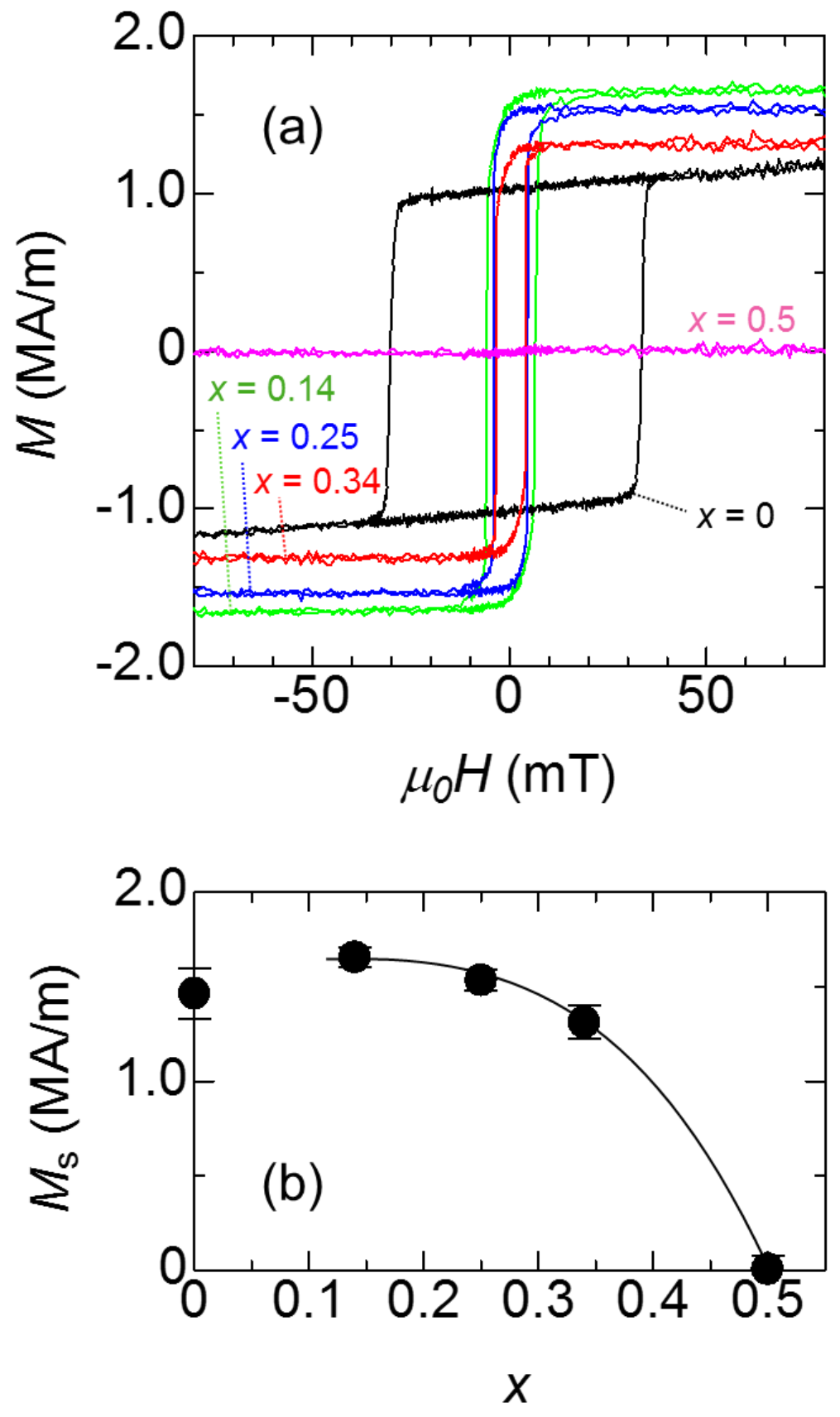

FIG. 3. (a) In-plane magnetization curves for $Co_{1-x}Mn_x$ films. (b) Mn concentration $x$ dependence of the saturation magnetization $M_s$. Note that the magnetization value for the Co film was obtained at saturation under an applied magnetic field of approximately 0.4 T. Curve in (b) is a visual guide.

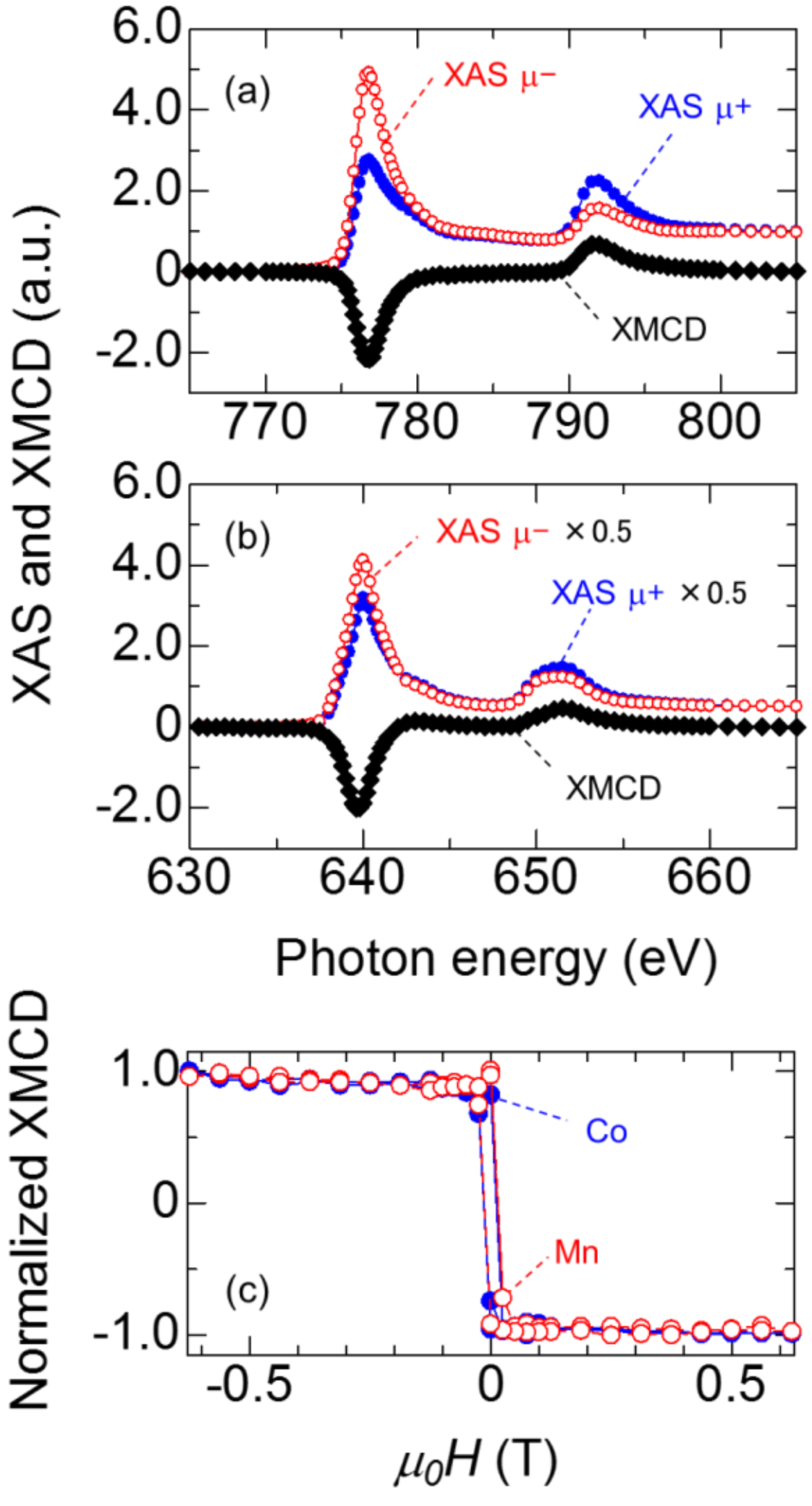

FIG. 4. XAS and XMCD of (a) Co and (b) Mn *L*-edges in the $x$ = 0.75 films. (c) Magnetic hysteresis curve of XMCD for the Co and Mn $L_3$-edge. The data are normalized.

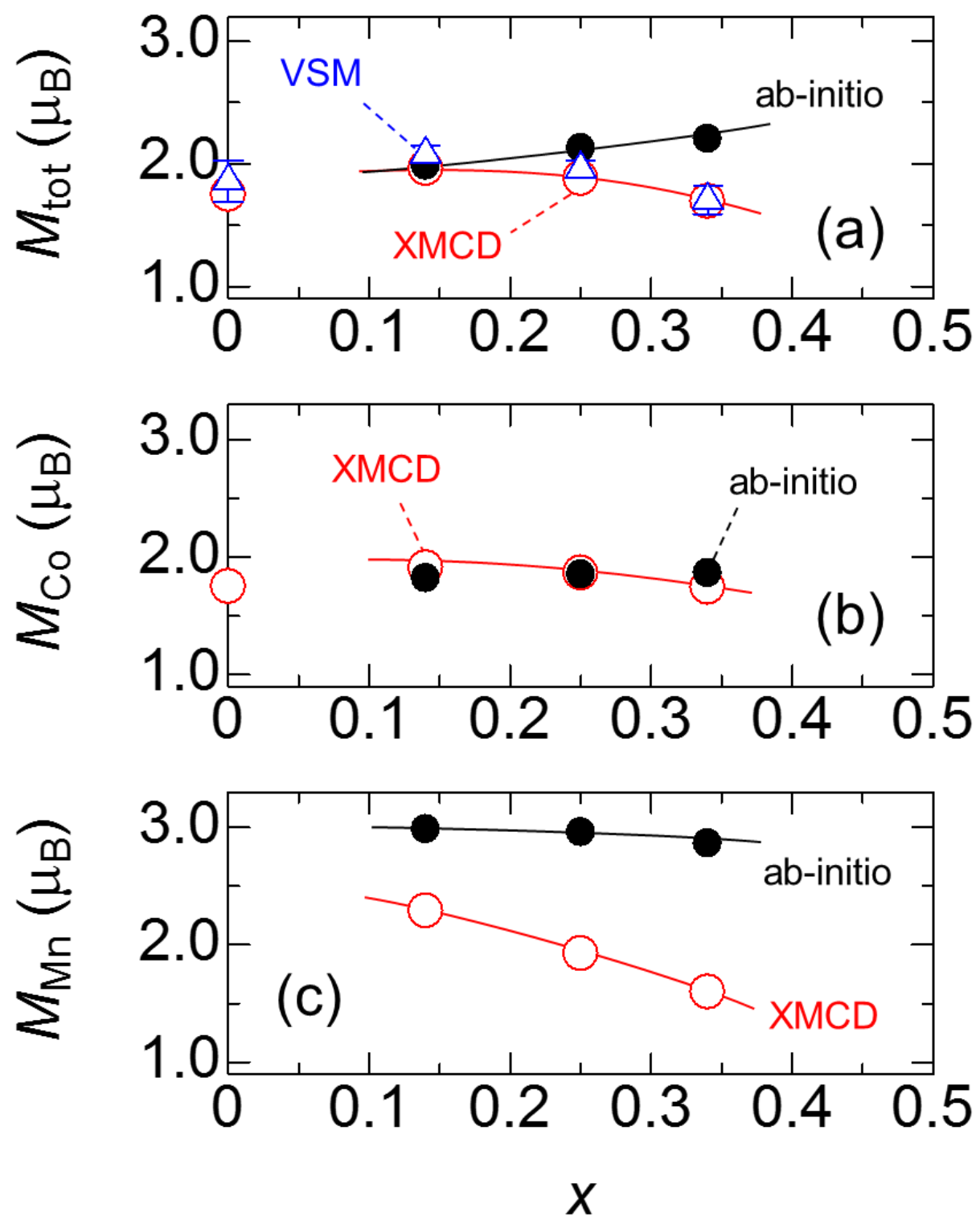

FIG. 5. (a) Total magnetic moment $M_{tot}$ obtained from the magnetization measurement, the sum rule of XMCD, and the *ab-initio* calculation. The element specific magnetic moment obtained from the sum rule of XMCD and *ab-initio* calculation for (b) Co $M_{Co}$ and (b) Mn $M_{Mn}$. Curves are visual guides.

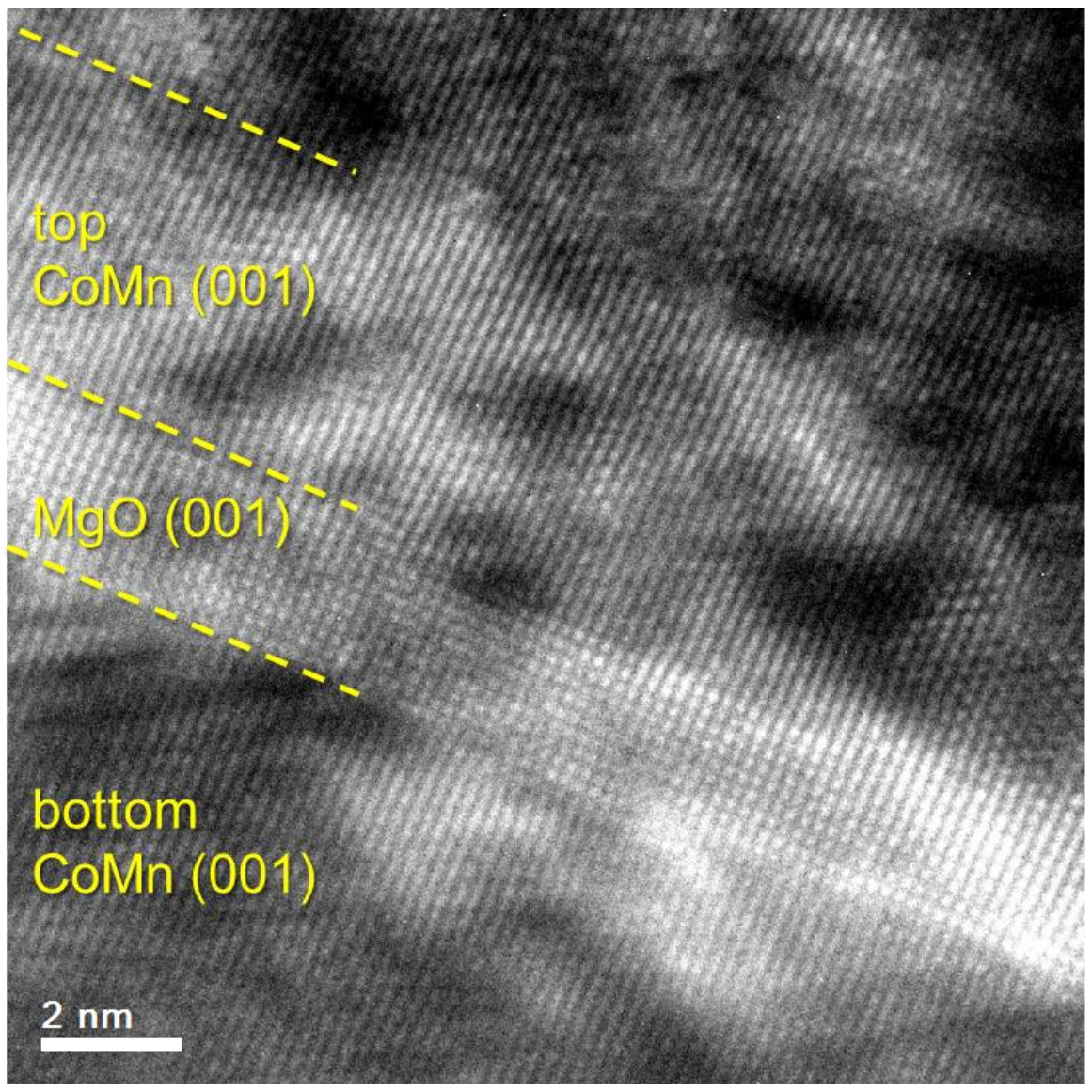

FIG. 6. The cross-sectional TEM image for the MTJ with the electrodes of the $x = 0.34$ films.

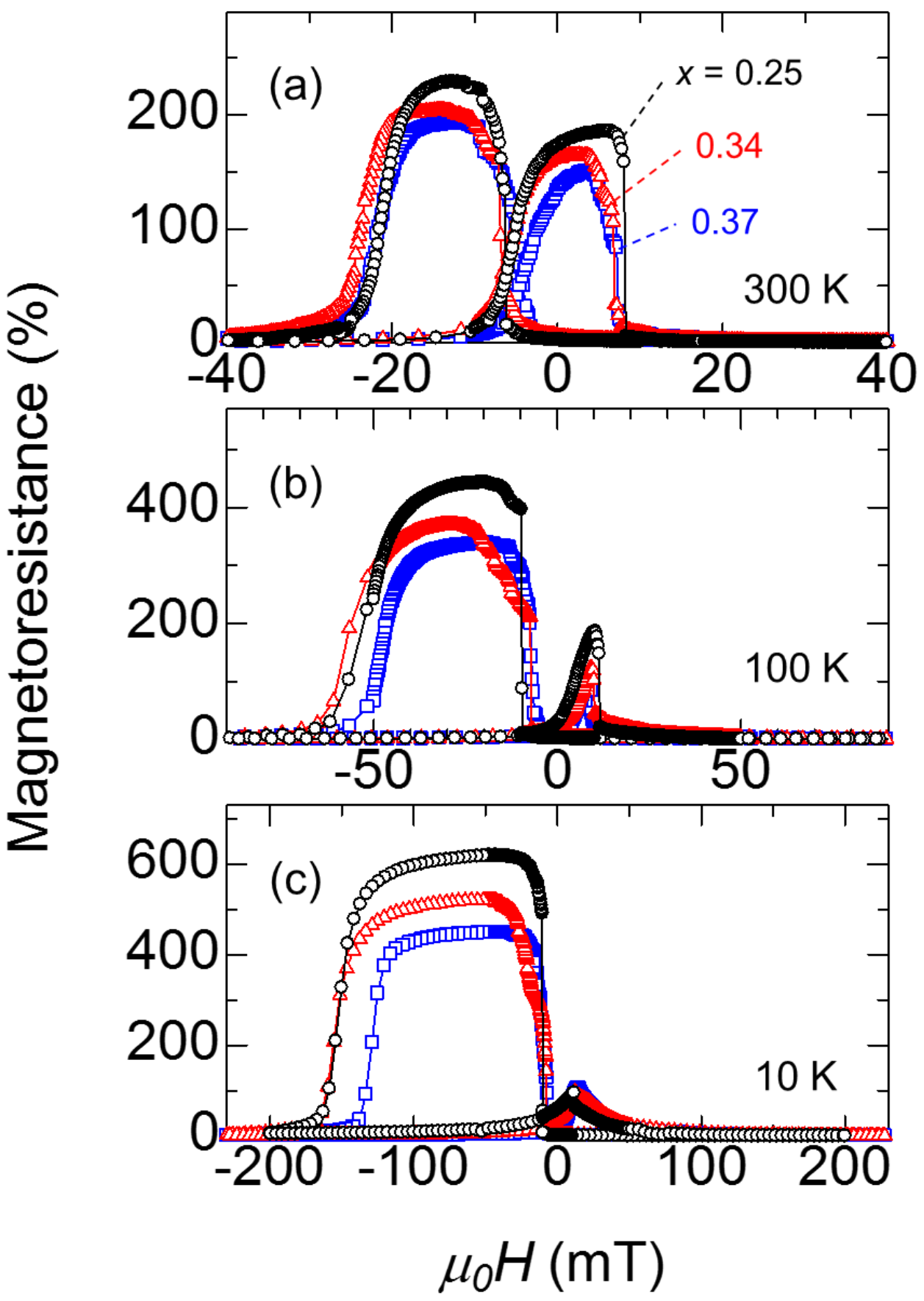

FIG. 7. The MR curves for the MTJs with the $Co_{1-x}Mn_x$ electrodes with $x$ = 0.25, 0.34, and 0.37, measured at (a) 300 K, (b) 100 K, and (c) 10 K.

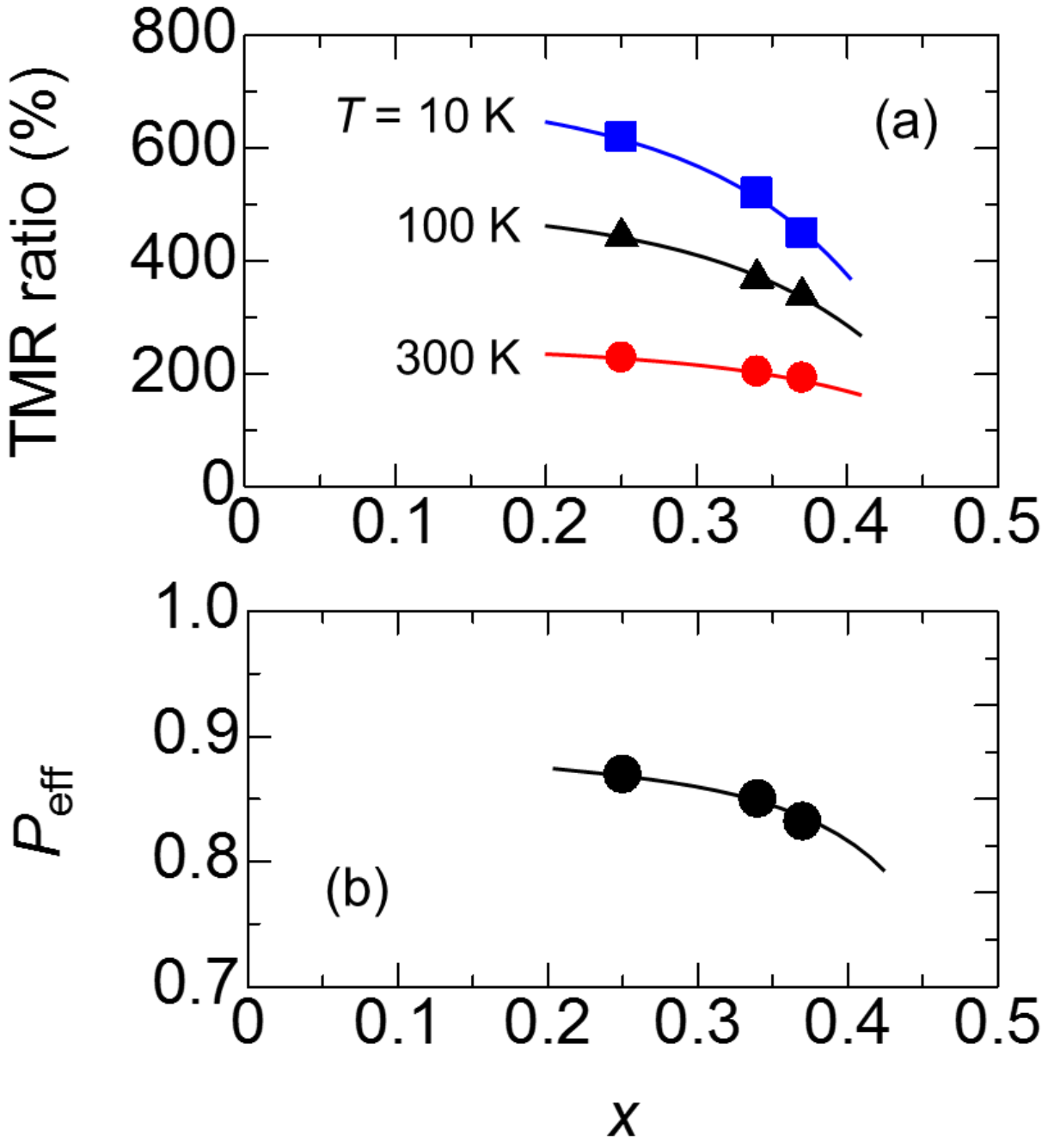

FIG. 8. (a) TMR ratios as a function of the Mn concentration *x* with different temperatures. (b) The Mn concentration *x* dependence of the effective spin polarization evaluated from the TMR ratio at 10 K with the Julliere's model. Curves are guides for eyes.

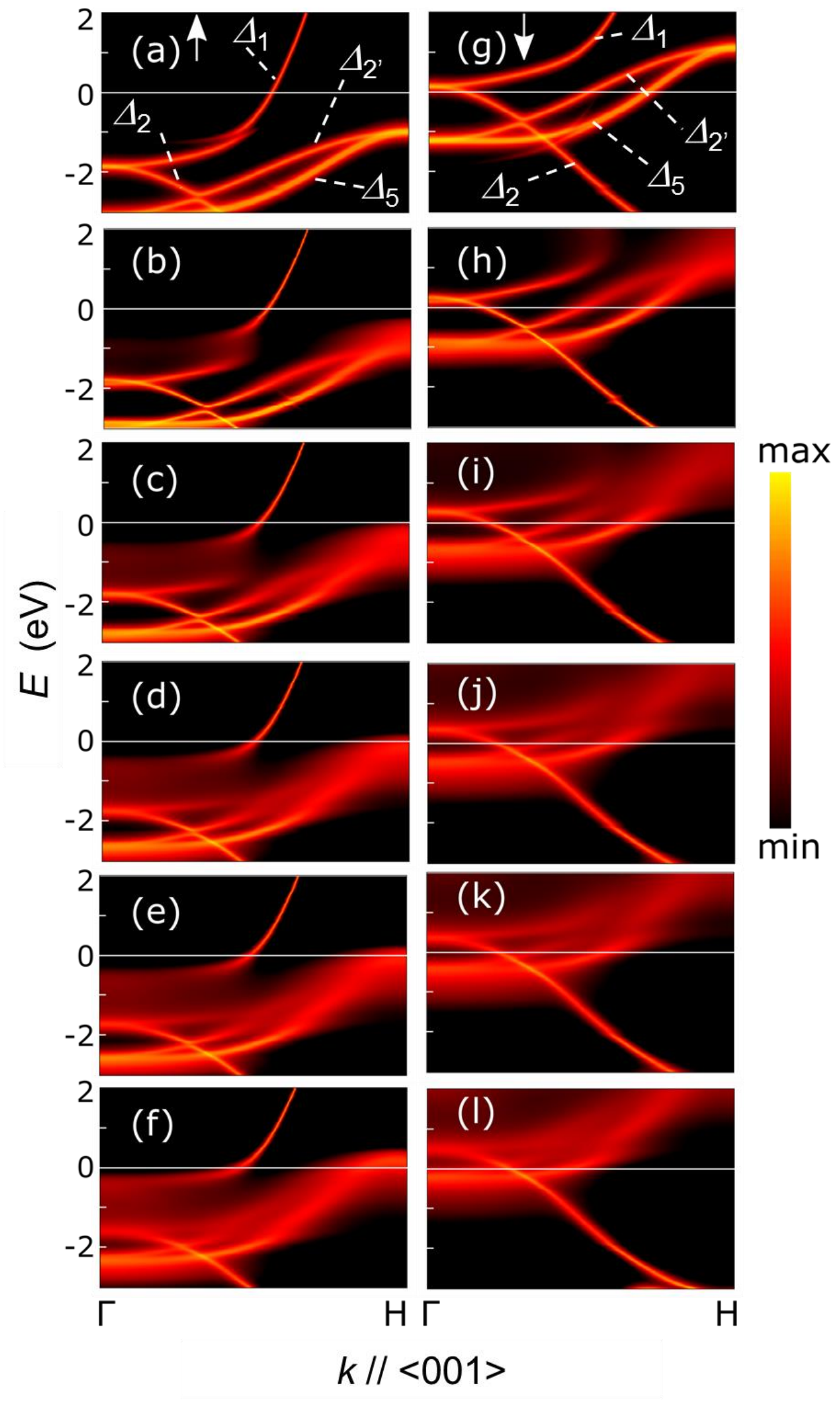

FIG. 9. The spin-resolved energy bands for bulk *bcc* $Co_{1-x}Mn_x$ calculated. (a) [(g)] $x = 0$, (b) [(h)] $x = 0.14$, (c) [(i)] $x = 0.25$, (d) [(j)] $x = 0.34$, (e) [(k)] $x = 0.37$, (f) [(l)] $x = 0.50$ for majority (minority) spin states. The lattice constant for *bcc* Co is assumed to be 0.282 nm, and the experimental values are used for other $x$.

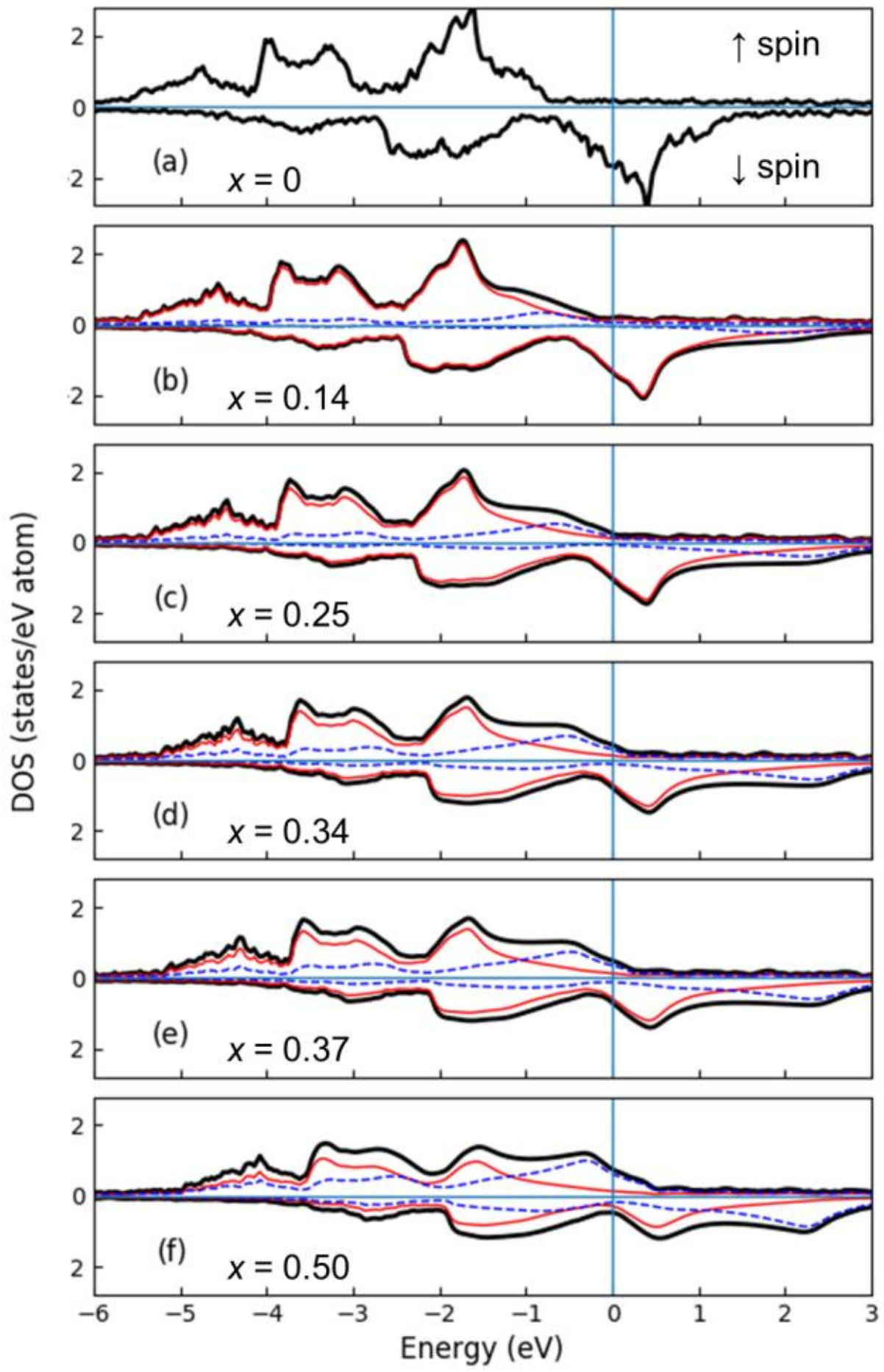

FIG. 10. The spin-resolved total and partial density of states for bulk *bcc* $Co_{1-x}Mn_x$ calculated. (a) $x = 0$, (b) $x = 0.14$, (c) $x = 0.25$, (d) $x = 0.34$, (e) $x = 0.37$, (f) $x = 0.50$. Thick (black), thin (red), and dashed (blue) curves correspond to the total density of states and partial density of states of Co and Mn atoms, respectively. The lattice constant for *bcc* Co is assumed to be 0.282 nm, and the experimental values are used for other *x*.